\documentclass[10pt, conference]{IEEEtran}
\IEEEoverridecommandlockouts

\usepackage{amsmath,amssymb,amsfonts}
\usepackage{algorithmic}
\usepackage{cite}
\usepackage{graphicx}
\usepackage{textcomp}
\usepackage{subcaption}
\usepackage[svgnames]{xcolor}
\usepackage{multirow}
\usepackage[hidelinks]{hyperref}
\usepackage{makecell}
\usepackage{booktabs}
\usepackage{listings}
\usepackage{tikz}
\usepackage{pifont}
\usepackage{comment}
\usepackage{enumitem}
\usepackage[most]{tcolorbox}
\usepackage{scalerel}
\usepackage{makecell}
\usepackage{caption}
\usepackage{float}
\usepackage{tcolorbox}
\tcbuselibrary{listings, skins, breakable}

\usepackage{longtable}
\usepackage{arydshln}

\usepackage[all=normal,paragraphs=tight,floats=tight,mathspacing=tight,wordspacing=tight,leading=tight,mathdisplays=tight]{savetrees}

\tcbset{colback=blue!1!white,colframe=blue!60!black,fontupper=\small,breakable,size=small}

\newcommand{\ourwork}{{SyntheticHLS}}

\title{\Huge\ourwork: Building Diverse Synthetic\\High-Level Synthesis Datasets using LLMs}

\author{
Stefan Abi-Karam\textsuperscript{1,2,*},
Miaoyan Zhou\textsuperscript{1,*},
Callie Hao\textsuperscript{1}
\\
Georgia Institute of Technology\textsuperscript{1},
Georgia Tech Research Institute\textsuperscript{2}
\\
\{\href{mailto:stefanabikaram@gatech.edu}{\nolinkurl{stefanabikaram}},
\href{mailto:mzhou346@gatech.edu}{\nolinkurl{mzhou346}},
\href{mailto:callie.hao@gatech.edu}{\nolinkurl{callie.hao}}\}@gatech.edu
\thanks{\textsuperscript{*}These authors contributed equally to this work.}
}

\begin{document}
\bstctlcite{IEEEexample:BSTcontrol}

\maketitle

\begin{abstract}

Deep learning and large language models (LLMs) have seen rapid adoption for semiconductor design, driving demand for hardware design datasets for model training. While most dataset efforts focus on hardware description languages (HDLs), high-level synthesis (HLS) remains a popular approach for domain-specific accelerators. Yet HLS designs are far scarcer. Recent HLS dataset efforts emphasize manual curation or design parameterization, with little effort toward generating high-quality synthetic HLS datasets using LLMs.

However, creating synthetic HLS datasets using LLMs presents significant challenges, including enforcing diversity across dimensions such as code length, design hierarchy, design space size, latency, resource utilization, and target application domain. We find that prior HLS dataset works seldom address this design diversity issue, leading to limitations in model generalization if used for deep learning model training.

Therefore, we propose \ourwork{}, a framework for generating large-scale, complex, and diverse synthetic HLS design datasets using LLMs. We present two key ideas: 1) an iterative feedback-guided mutation loop that incrementally transforms seed designs into more complex and "well-scalable" final designs using paired HLS source code and design-space specifications, and 2) quantitative metrics for HLS design complexity and design-space scalability that provide measurable target objectives for LLM-guided mutation.

Lastly, we conduct a systematic cross-validation of an HLS Quality-of-Results (QoR) deep learning model trained and tested across common HLS benchmarks, zero-shot synthetic designs, and iteratively mutated synthetic designs. Our results show that synthetic designs transfer well to common benchmark test sets while the reverse does not hold, and that iteratively mutated designs from \ourwork{} provide the most generalizable training corpus among the datasets studied. We further characterize the mutation process and resulting synthetic dataset, showing that metric-guided trajectories consistently improve targeted complexity and scalability objectives without regressing non-target metrics, and that mutated designs span a substantially broader and more diverse design space than zero-shot generated designs.

We open-source our framework, dataset, and evaluation at \textcolor{blue}{\url{https://github.com/sharc-lab/synthetic-hls}}.

\end{abstract}

\section{Introduction}

High-level synthesis (HLS) has become a dominant methodology to build domain-specific accelerators for real-time and compute-intensive applications (e.g., autonomous navigation \cite{Wan2022RoboticCO}, high-energy physics \cite{dgnnflow, duarte2018fasta, kvapil2025intelligent}, AI inference \cite{scalehls, allo, chen2024understanding}). However, achieving high-quality and performant hardware from an HLS design requires navigating a large and complex design space. Users can specify compiler directives, i.e., "pragmas",  to apply optimizations such as loop unrolling, array partitioning, and pipelining among others \cite{hls_expert_design}. This complexity has driven substantial prior work applying machine learning (ML) and deep learning to quality-of-result (QoR) prediction and design space exploration (DSE) for HLS design \cite{dai2018fast, wu2021ironman, wu2022ironman, harp, hlsyn, autodse, lift, sohrabizadeh2022automated}. 

A core challenge of these deep learning methods is the dataset diversity of HLS designs used for training. Ideally, models should generalize across a wide range of input HLS designs, which requires diversity along two axes: 1) \emph{configuration diversity}, sampling many designs with different pragma configurations from a single base design and 2) \emph{design diversity}, covering many distinct HLS designs from different application domains. Prior works focus predominantly on the first axis, sampling hundreds of configurations from a small set of common benchmark designs \cite{hlsyn}. This leaves model generalization across diverse and unseen HLS designs a challenge. Manual and semi-automated curation of published open-source HLS designs offers one path forward \cite{abi-karam_hlsfactory_2024}. However, we propose an alternative: to generate synthetic HLS design data at scale using large language models (LLMs).

Synthetic HLS design generation remains sparsely explored \cite{iceberg}, leading to many open problems for this approach. Zero-shot LLM generation is a natural first step, but the resulting designs are small, simple, and unrepresentative of real-world HLS designs which are typically parameterizable across a wide range of latency--resource trade-off points. We argue that this \textbf{lack of complexity and diversity} will cause deep learning models trained on \textbf{naively LLM-generated synthetic designs} to \textbf{generalize poorly} on HLS design tasks. Furthermore, there is a \textbf{lack of systematic cross-validation analysis} of how well deep learning HLS QoR models generalize across design datasets in general. Without addressing these gaps, LLM-generated synthetic HLS designs offer limited practical value.

Therefore, we propose \textbf{\ourwork{}}, a comprehensive framework for generating large-scale, complex, diverse, and scalable synthetic HLS design datasets. Our contributions are as follows:

\begin{figure*}[th!]
    \centering
    \includegraphics[width=\textwidth]{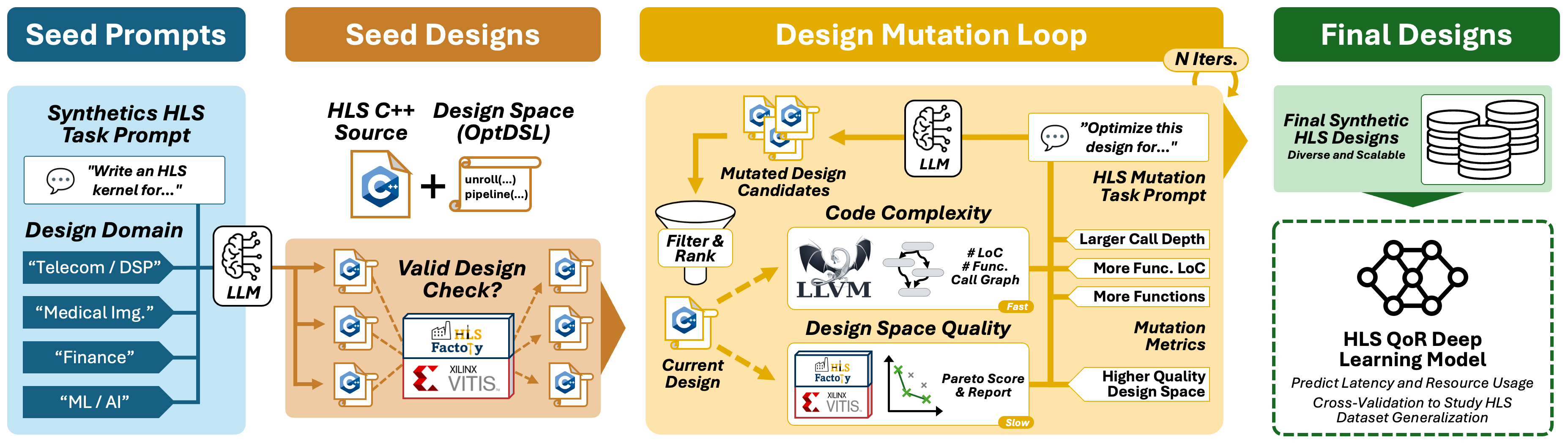}
    \caption{Overview of \ourwork{} highlighting our domain-driven seed design generation as well as our iterative metric-guided design mutation to create a final set of diverse, complex, and well-scalable HLS designs for generalizable deep learning training.}
    \label{fig:main}
    \vspace{-1.5em}
\end{figure*}

\setlist[itemize]{leftmargin=1em}
\begin{itemize}
    \item \textbf{Iterative Feedback-Guided Mutation}: We present an iterative design mutation loop that pushes designs to be more diverse and scalable through LLM-guided design editing to optimize selected design metrics at each stage. This results in high-quality designs beyond just naive zero-shot design generation.

    \item \textbf{Quantifying HLS Design Complexity and Scalability}: We introduce quantitative metrics that score candidate designs on structural complexity and design-space scalability; these metrics provide measurable objectives for LLM-guided design mutation.

    \pagebreak
    
    \item \textbf{Cross-Validation of Synthetic and Benchmark Designs for QoR Modeling}: We conduct the first systematic cross-validation of HLS QoR deep learning modeling across synthetic and benchmark datasets, showing that iteratively mutated \ourwork{} designs generalize better than zero-shot synthetic or common benchmark designs alone.

    \item \textbf{Trajectory and Dataset Characterization}: We characterize both the mutation process and its resulting dataset, showing that metric-guided trajectories consistently improve targeted objectives without regressing non-target metrics, and that mutated designs occupy a substantially broader and more diverse design space than zero-shot generated designs.
    
    \item \textbf{Open-Source Scalable Framework}: We release \ourwork{} as an open-source Python package with a parallel execution backend, enabling others to run efficient large-scale synthetic HLS dataset generation.    
\end{itemize}

\section{Prior Work}

Several HLS benchmarks and datasets are widely used for ML and non-ML HLS research, including PolyBench \cite{polybench}, MachSuite \cite{machsuite}, CHStone \cite{chstone}, Rosetta \cite{rosetta}, HLSyn \cite{hlsyn}, and HLSFactory \cite{abi-karam_hlsfactory_2024}. Some are standalone curated collections of HLS designs (e.g., PolyBench), while others aggregate and sample these base benchmarks to produce more parameterized design points (e.g., HLSFactory, HLSyn). In both cases, generalization is a challenge when using these datasets for deep learning training. Source benchmarks follow fixed computational patterns for a narrow domain, while aggregation-based datasets enumerate thousands of points from a small set of source designs. Training a QoR proxy model on these design datasets will not generalize well to unseen designs. Although these datasets have been valuable for building proxy modeling and DSE approaches, producing more diverse and generalizable HLS training data remains an open problem.

The closest prior work on synthetic HLS design generation is Iceberg \cite{iceberg}, which also explores LLM-generated synthetic HLS designs. However, Iceberg only presents a zero-shot setting where an LLM generates only design source code. The design space is also produced separately via static analysis rather than by the LLM. They apply some basic prompting techniques to encourage application diversity and nudge designs toward target loop counts or memory use, but all still in a zero-shot setting, i.e., ask the model for a design and use the first output. These two limitations, zero-shot generation and non-LLM design-space construction, restrict the diversity, complexity, and scalability of the resulting designs, weakening their utility for QoR model training. In general, Iceberg explores synthetic data generation as an exploratory component of the wider work rather than a systematic study of how to generate high-quality synthetic HLS designs.

Finally, we find the generalization gap in HLS QoR modeling remains understudied. Existing works train and evaluate on large datasets sampled from only a few source designs. No systematic cross-validation exists across different training and testing datasets, whether drawn from common benchmarks or synthetic designs, to identify which sources or mixtures generalize best. Therefore, we also aim to present a strong set of initial generalization results for HLS QoR proxy models in our work.

\section{\ourwork{} Framework}

To tackle LLM-guided synthetic HLS design generation, we present our framework \ourwork{} in Fig.~\ref{fig:main}. Design generation is broken into two high-level phases: seed design generation and iterative design mutation. The designs collected after iterative design mutation have been optimized for diversity, code complexity, and Pareto frontier scaling, making them high-quality training data for deep learning training.

\subsection{\textbf{Seed Design Generation}}

In the first stage, \ourwork{} uses an LLM to generate a diverse set of seed designs. Each seed includes source code for an HLS kernel, a self-contained testbench file, and a design space specification using our OptDSL syntax (discussed further in \S\ref{sec:pareto}). These designs are then filtered to retain only compilable, functional, and synthesizable designs as the final seed design set.

To ensure seed designs reflect realistic, diverse workloads, we guide LLMs to produce application-oriented kernels across multiple domains rather than isolated micro-kernels. Each kernel is conditioned on a user-selected domain description with corresponding applications adopted from Iceberg~\cite{iceberg}. We also instruct the LLM to decompose each design into multiple interacting stages with explicit dataflow, memory, and control dependencies. This structure yields seed designs with non-trivial program structure and varied memory patterns across domains. As a result, the seed design generation stage in \ourwork{} is comparable to the approach used to generate designs in Iceberg. In our later results we denote these seed designs as \textit{\textbf{Base}} designs.

\subsection{\textbf{Iterative Design Mutation}}

In the second stage, \ourwork{} iteratively improves each seed design by optimizing a sequence of target metrics. Each iteration generates a batch of candidate designs, evaluates them for correctness and synthesizability, and selects the best candidate as the seed for the next iteration. This loop follows a user-specified schedule of optimization targets and iteration counts, ultimately producing a curated set of passing synthetic HLS designs with associated evaluation artifacts for downstream ML and EDA tasks. In our later results we denote the set of designs after iterative mutation as \textit{\textbf{Final}} designs.

\subsubsection{\textbf{Feedback-Guided Mutation}}
At each iteration, \ourwork{} constructs a mutation prompt from three inputs: 1) the current design's files, 2) the target metric for the current iteration, and 3) a structured evaluation report of the current design. This report includes one or both of the following artifacts:
\texttt{call\_graph.json}, a structural summary of the design's source code, and \texttt{pareto\_scores\_summary.json}, a design-space scalability report from Pareto-score analysis. We offer more discussion in \S\ref{sec:metrics} on these target metric data.

Multiple candidate output designs are generated from an LLM using this single mutation prompt through repeated LLM sampling. One of these candidate mutated designs is then selected to be the input into the next iteration according to the candidate selection outlined in the following section.

The number of candidate samples generated per design and the schedule of design mutation (i.e., the number of design iterations, what metric is targeted on each iteration) is user-definable, allowing the end user to explore LLM inference scaling techniques across different LLMs (highlighting exploratory future work for \ourwork{}).

\subsubsection{\textbf{Candidate Selection}}

\begin{figure}[t!]
    \centering
    \includegraphics[width=0.5\textwidth]{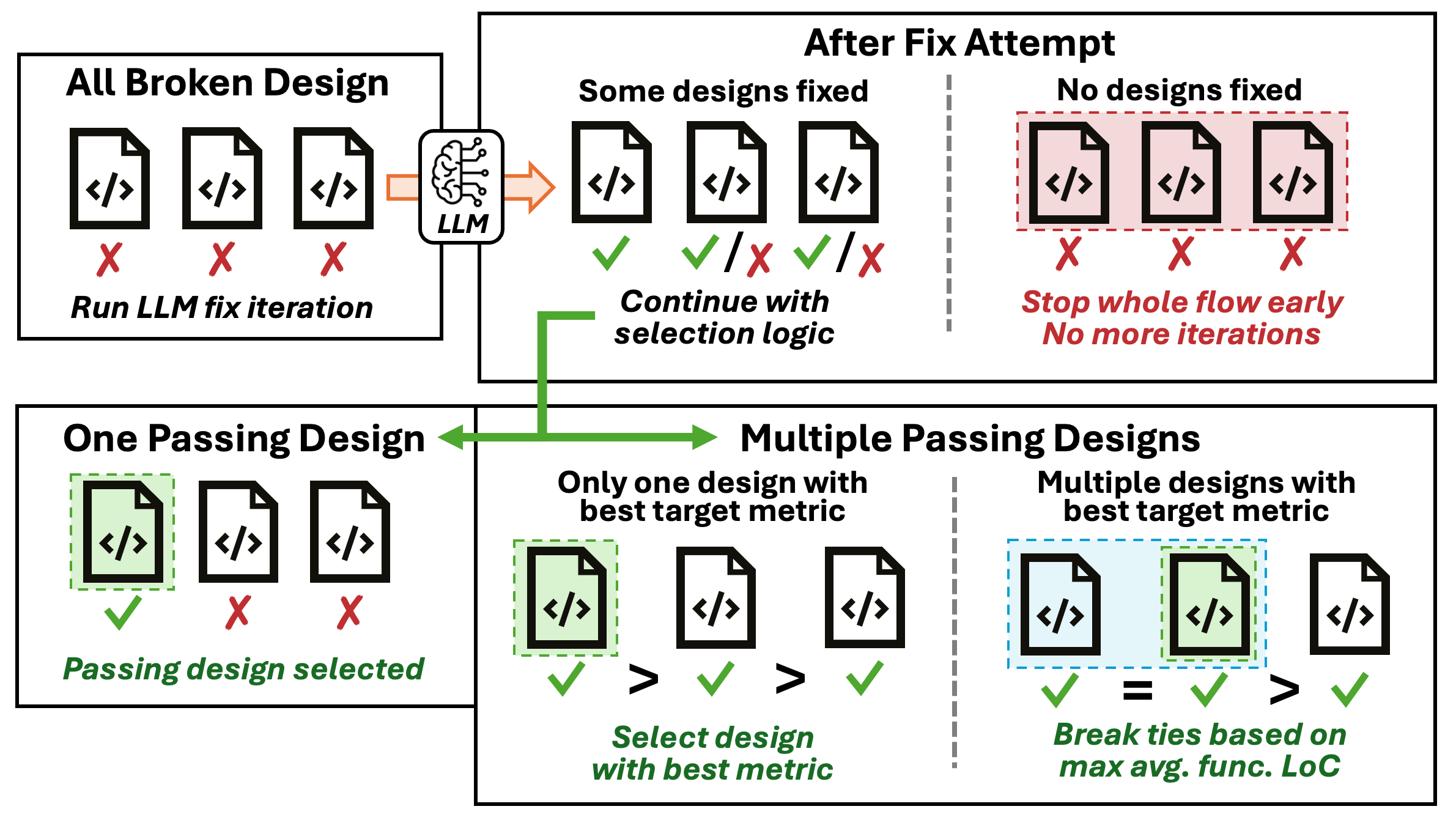}
    \caption{Selection strategy for picking the final design at each mutation stage. For passing designs, select the candidate with the best target metric, breaking ties with avg. func. LoC. If all candidates fail, attempt LLM-based repair; if none are fixable, the trajectory is terminated.}
    \label{fig:selection}
\end{figure}

After each iteration, SyntheticHLS selects the next-iteration design from the current evaluated candidate design batch using the selection strategy summarized in Fig.~\ref{fig:selection}. A mutated candidate is considered valid if it compiles and synthesizes successfully, verified by running the \texttt{csim} and \texttt{csynth} flows of Vitis HLS. Based on the mix of passing and failing candidates and their target-metric values, SyntheticHLS applies the following selection logic to determine the picked candidate for the next iteration.

\textbf{One passing candidate:}
The single passing design is selected directly as the next-iteration seed.

\textbf{Multiple passing candidates, unique best:}
All passing candidates are ranked by their target-metric value. The candidate achieving
the highest value is selected as the next-iteration seed.

\textbf{Multiple passing candidates, tied best:}
If several passing candidates share the maximum target-metric value, \ourwork{} breaks the tie by picking the tied design with the max average function LoC metric.

\textbf{No passing candidates:}
If the entire candidate batch fails evaluation, \ourwork{} triggers a "fix" phase. For each failed design, a repair prompt comprising the complete failed design, the compiler or HLS synthesis error messages, and fix-oriented instructions is constructed, prompting the LLM to fix the design while preserving functional behavior. The repaired designs are re-evaluated, and the other selection rules are applied to any designs that now pass. If no repaired design passes, the entire iteration flow for the current design terminates early.

Because our selection logic repairs candidates only when all fail, \ourwork{} may bias which mutated designs advance to the next stage toward less diverse or complex designs, which are more likely to be generated correctly by the LLM. However, we view this as a deliberate trade-off between bias and compute/runtime: repairing every failed candidate would roughly double LLM and synthesis invocations. Whether universally repaired candidates increase corpus diversity is an interesting question we leave to future work. Nonetheless, our later results characterize design failure rates at different mutation stages, offering some insight into the behavior of this selection logic.

\subsection{\textbf{Quantifying HLS Designs}}

\label{sec:metrics}

To meaningfully mutate and diversify LLM-generated HLS designs, we need quantitative metrics that, when optimized, increase design complexity and diversity. We therefore introduce two types of design metrics: \textbf{structural complexity metrics}, describing the hierarchical and modular organization of the HLS source code, and \textbf{design-space metrics}, describing how scalable and parameterizable a design is. Together, these provide measurable, actionable objectives that guide the LLM toward targeted improvements in each iteration of the mutation loop.

\subsubsection{\textbf{Structural Complexity}}
Structural code complexity metrics let us identify "naive" HLS designs and use LLMs to mutate them to be larger and more complex. Zero-shot LLM generation produces small, simple kernels (see design metrics at the starting iteration in Fig.~\ref{fig:trj}). By iteratively prompting the LLM to rewrite a design for higher structural complexity, the final design becomes "larger" (more compute/resources) and "more complex" (more control flow: loops + conditionals).

We quantify the structural complexity of HLS kernels through static analysis of each kernel's C/C++ source code. Using Clang's AST dump feature, we extract function declarations, construct a directed call graph, and compute three metrics:
\begin{itemize}
    \item \textbf{Number of Functions}: Distinct function count, reflecting modular decomposition.
    \item \textbf{Maximum Call-Chain Depth}: Longest call path, reflecting hierarchical depth.
    \item \textbf{Average Lines-of-Code (LoC) per Function}: Typical function size, reflecting whether computation is grouped in larger or smaller units.
\end{itemize}

Together, these metrics tell the LLM how much functionality is factored into reusable units, how deeply computation is nested, and how complexity is distributed across the design.

\definecolor{hlsbg}{RGB}{242,242,235}
\definecolor{hlsborder}{RGB}{210,210,200}

\lstdefinestyle{codefig}{
  basicstyle=\ttfamily\scriptsize,
  columns=fixed,
  keepspaces=true,
  showstringspaces=false,
  aboveskip=0pt,
  belowskip=0pt,
  breaklines=true,
  backgroundcolor=\color{hlsbg},
  frame=single,
  frameround=f,        
  rulecolor=\color{hlsborder},
  framesep=4pt,           
  framerule=0.5pt,        
  xleftmargin=4pt,        
  xrightmargin=4pt,
}

\begin{figure}[t!]

  \centering

  \lstset{style=codefig}
  \begin{lstlisting}
partition("A", "k2mm", "cyclic", [1,2,4,8], 1, "group1")
partition("A", "k2mm", "cyclic", [1,2,4,8], 1)
unroll("lp1", "k2mm", [1,2,4,8], "group1")
unroll("lp2", "k2mm", [1,2,4])
pipeline("lp1", "k2mm", optional=True)
pipeline("lp2", "k2mm")\end{lstlisting}
  
  \caption{OptDSL syntax for specifying HLS design spaces across HLS pragma types and pragma parameter values.}
  \label{fig:OptDSLv2}
\end{figure}

\subsubsection{\textbf{Pareto Score}}

\label{sec:pareto}

A complementary way to evaluate an HLS design is by the size and quality of its design space. Each HLS source file is paired with a design-space specification that enumerates possible pragma values; sampling this space yields thousands of variants of a single design.

A \textit{well-scalable} design has a large space whose Pareto frontier spans many points across a wide latency–resource range. Because zero-shot LLM generation tends to produce poorly scalable designs, we introduce a metric that guides the LLM toward better-scalable source code and design-space specifications. Optimizing this metric encourages the LLM to mutate \textbf{both} the source code and the pragma configuration, yielding a richer, more scalable design.

To quantify the scalability of a design, we introduce our \textbf{Pareto Score} metric inspired by $\Delta$-index metrics from multi-objective optimization~\cite{opt_metrics, ngsa_ii}. This score captures how effectively a design spans the latency--resource trade-off frontier, reflecting design-space size and diversity.

Let $x$ denote a resource metric (e.g., LUT or FF counts) and $y$ denote latency (clock cycles). After normalization to user-specified scales or observed ranges, the Pareto frontier $P=\{(\tilde{x}_i,\tilde{y}_i)\}$ is extracted and sorted by $\tilde{x}$. The score is then defined as:
\[
S_P = \frac{d_{\text{extent}} + G_{\max}}{d_{\text{extent}} + L}.
\]
\\
Here, $L=\sum_i d_i$ is the total frontier length, where $d_i = \big\|(\tilde{x}_{i+1},\tilde{y}_{i+1}) - (\tilde{x}_i,\tilde{y}_i)\big\|_2$ is the Euclidean distance between consecutive points. The largest gap, $G_{\max} = \max_i d_i$, penalizes discontinuous frontier coverage. The term $d_{\text{extent}}$ is the summed distances from the frontier endpoints to the two ideal corners $(0,1)$ and $(1,0)$, corresponding to the best possible low-latency and low-resource implementations, respectively.

By construction, $S_P \in (0,1]$, where lower scores indicate a smoother, denser, more continuous frontier. When this metric is the optimization target for the current iteration, we ask the LLM to produce a mutated design and design-space specification that minimizes $S_P$.

\textbf{Design Space Specification}: Computing the Pareto score requires sampling and evaluating design configurations via HLS synthesis to construct a design space, extract a Pareto frontier, and calculate the score.

Each design maintains a paired design space throughout its trajectory. We use an improved version of OptDSL, the design space DSL from HLSFactory \cite{abi-karam_hlsfactory_2024}, which supports array partitioning, loop unrolling, and pipelining pragmas with parameterizable and grouped pragma parameters. We highlight an example OptDSL design space in Fig.~\ref{fig:OptDSLv2}.

For each candidate evaluation, we randomly sample $N$ design points (64 by default to keep synthesis runtime tractable). We run the full HLS synthesis flow on each point to obtain QoR metrics, then compute Pareto frontiers for LUTs vs. latency and FFs vs. latency, scoring each with our Pareto-score metric. The resulting summaries are written to \texttt{pareto\_scores\_summary.json} and fed into the iterative mutation loop.

\subsection{\textbf{Extensible API and Parallel Backend}}

\ourwork{} is implemented as an open-source, extensible Python package. Users can configure synthetic HLS design generation runs to target different domains or tune scaling parameters such as the number of seed designs, the mutation schedule, and the number of candidates generated per mutation iteration.

To generate and evaluate large volumes of HLS designs efficiently, \ourwork{} uses a three-level parallel execution strategy. At the \textit{design level}, a shared thread pool manager instantiates three fixed-size pools, one each for LLM inference, C simulation, and HLS synthesis, bounding concurrent executions per tool type. At the \textit{run level}, a separate thread pool scales across LLM backends and application domains, concurrently executing seed generation and the iterative mutation loop. A user-specified job budget is split automatically across active runs to maximize throughput while keeping overall concurrency bounded. Finally, \textit{Pareto score computation} triggers a third level of parallelism by invoking HLSFactory to sample candidate design points and dispatch additional HLS tool calls in parallel.

\begin{figure*}[ht!]
    \centering
    \includegraphics[width=\textwidth]{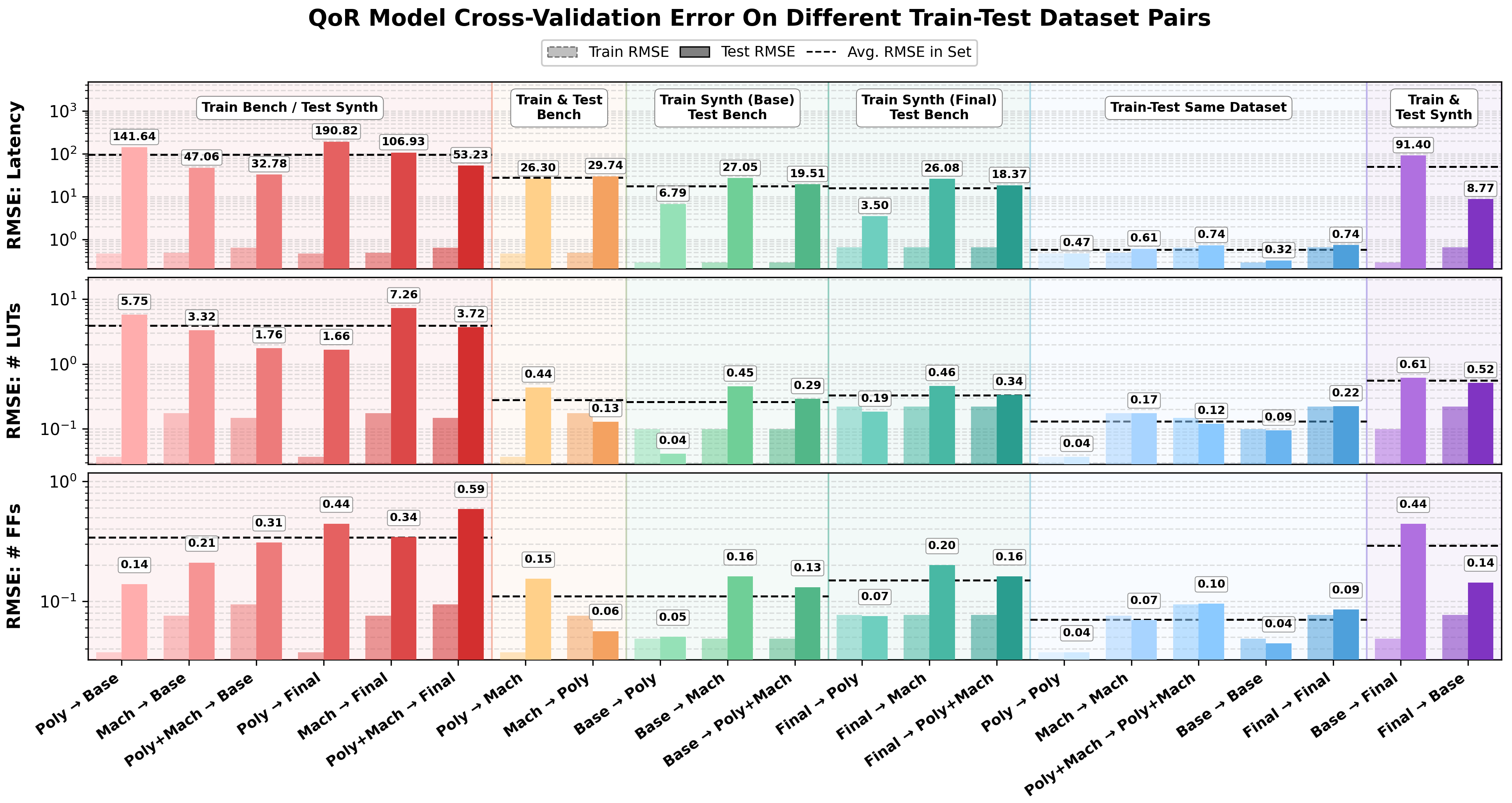}
    \caption{Cross-dataset generalization results for HLS QoR modeling. All bars report root-mean-squared error (RMSE) of the predicted value (lower is better). Each pair of bars denotes a \textbf{\texttt{[train]→[test]}} configuration, where training and test sets are drawn from: \textbf{\textit{Poly}} (PolyBench), \textbf{\textit{Mach}} (MachSuite), \textbf{\textit{Poly+Mach}} (their union), \textbf{\textit{Base}} (zero-shot synthetic), and \textbf{\textit{Final}} (feedback-mutated synthetic). Synthetic training data generalizes broadly to benchmark test sets (green region), while the reverse does not hold (red region).}
    \label{fig:results_bar}
    \vspace{-1.5em}
\end{figure*}

\begin{figure*}[t]
    \centering
    \includegraphics[width=\textwidth]{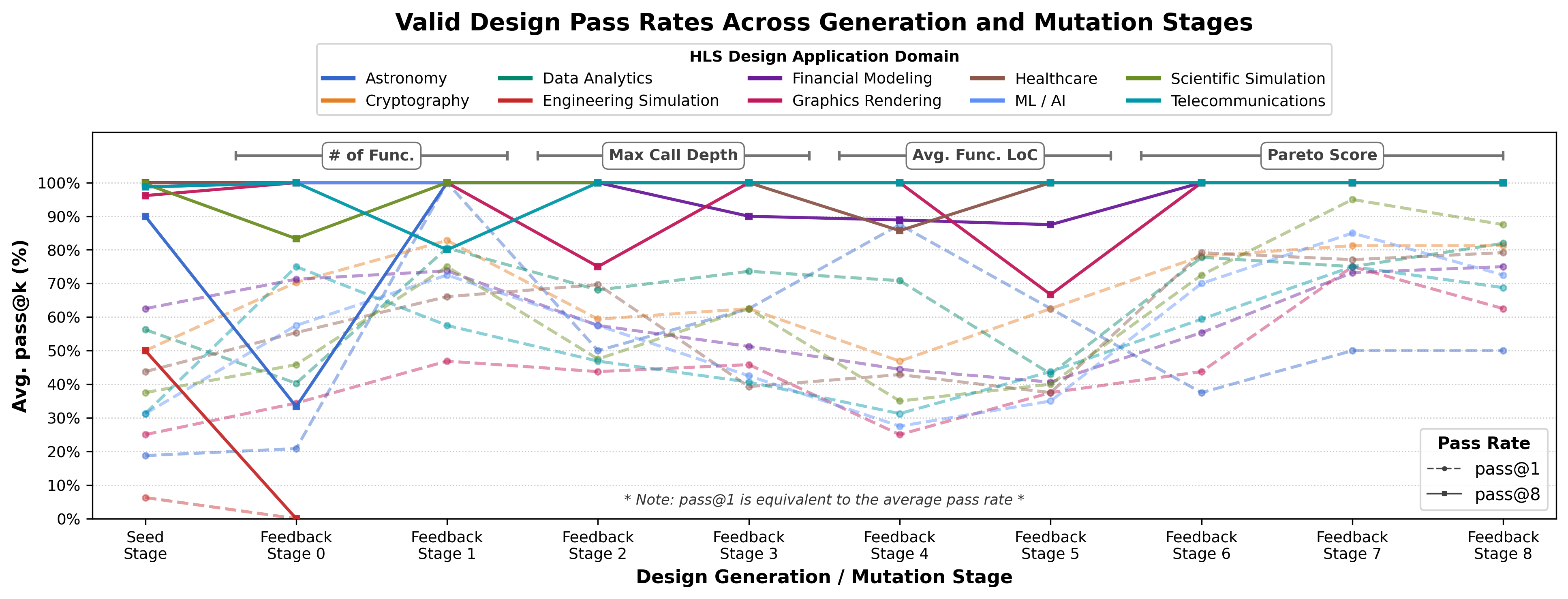}
    \caption{Average pass@1 and pass@8 rates of valid, synthesizable designs across the seed generation stage and nine feedback-driven mutation stages, broken down by HLS application domain. Each group of mutation stages targets a different design metric (number of functions, max call depth, average function LoC, and Pareto score). Pass@8 remains near $100\%$ for most domains and is $100\%$ at the final stage, while pass@1 ranges from roughly $30\%$ to $80\%$.}
    \label{fig:pass_rates}
\end{figure*}

\section{Evaluations}

To demonstrate the quality of \ourwork{} for generating generalizable synthetic HLS designs, we present two evaluations: 1) a cross-validation of a deep learning QoR proxy model trained on our synthetic designs and common HLS benchmarks, and 2) a demonstration of how design metrics improve over multiple iterations of a design trajectory.

\subsection{\textbf{Generalization of Synthetic HLS Designs for Deep Learning QoR Prediction}}

To evaluate the generalization of synthetic HLS designs, we construct a cross-validation experiment that trains and tests a QoR proxy model for predicting latency and resource usage, e.g., LUTs and FFs, of synthesized HLS designs only given the design's source code. By varying the train-test dataset pair while holding the model architecture fixed, we identify which training datasets generalize best across the most design sets.

We adopt the QoR modeling approach from GNN-DSE~\cite{sohrabizadeh2022automated}, which predicts latency and resource usage directly from HLS source code synthesized by Vitis HLS (in this evaluation we use v2023.1). Each HLS kernel's code is translated into a graph representation and fed into a graph neural network (GNN) with separate decoding heads for each target metric. Our training uses a batch size of 64, MSE loss, and 200 epochs. We defer to the GNN-DSE paper and our code for further implementation details.

We consider five design datasets as candidates for training and testing: \textbf{\textit{Poly}} (PolyBench benchmarks), \textbf{\textit{Mach}} (MachSuite benchmarks), \textbf{\textit{Poly+Mach}} (their union), \textbf{\textit{Base}} (zero-shot synthetic HLS designs from \ourwork{}'s seed design generation stage), and \textbf{\textit{Final}} (synthetic designs generated by \ourwork{} after feedback-guided LLM design mutation). Note that the \textbf{\textit{Base}} dataset serves as a proxy for the design generation method of the Iceberg prior work. Using consistent notation, we denote each cross-validation evaluation configuration as \texttt{[train\_name]$\to$[test\_name]}. For each dataset, we randomly sample 4,096 design points and apply a fixed 70\% / 30\% train-test split. We note that all cross-dataset configurations are inherently "out-of-distribution", since training and testing kernels are drawn from entirely different sources; conversely same-dataset evaluations evaluate training generalization "in-distribution".

The \textbf{\textit{Poly}}, \textbf{\textit{Mach}}, \textbf{\textit{Poly+Mach}}, \textbf{\textit{Base}}, and \textbf{\textit{Final}} datasets contain 28, 18, 46, 68, and 68 unique kernels, respectively. \textbf{\textit{Base}} and \textbf{\textit{Final}} span 10 domains: astronomy (6), cryptography (5), data processing (5), engineering simulation (3), financial modeling (14), graphics/rendering (3), healthcare/medical (7), ML/AI (11), scientific simulation (9), and telecommunications/DSP (5). Synthetic designs were generated with up to 32 seed designs per domain and 8 mutation candidates per iteration. The mutation schedule targets number of functions (2 iters.), call-chain depth (2 iters.), average function lines (2 iters.), and Pareto scores (3 iters.) sequentially, with Pareto scoring computed over 64 samples per candidate. The combined synthetic datasets were collected over approximately 24 hours of wall-clock time.

Although our selected configuration bounds the generated dataset to a specific size, \ourwork{} is not inherently fixed in scale. Unlike existing HLS datasets, which are typically fixed in size and parameterizable primarily through pragma configurations of fixed kernels, \ourwork{} can scale directly with additional LLM and synthesis compute, and also parameterizes designs across both program structure and pragma configurations.

All synthetic designs were generated using the \texttt{gpt-oss-120B} LLM. We deliberately select this model for three reasons: 1) as an open-weights model, it enables accessible reproducibility of our generation pipeline via low-cost cloud inference or free local inference for users with the hardware; 2) its permissive license avoids data-provenance restrictions on redistributing the generated dataset, which some proprietary model terms constrain; and 3) as a strong but non-frontier model, it provides a conservative lower bound on framework performance, i.e. if \ourwork{} produces high-quality designs with this weaker model, it can perform as well or better with frontier open- and closed-weights models. \ourwork{} is model-agnostic, and we view a systematic study across LLM families and scales as important future work.

\textbf{Results}: Our cross-evaluation results are comprehensively shown in Fig.~\ref{fig:results_bar}. The results presented offer several key insights into how training data composition shapes HLS QoR model generalization, a topic that has received little systematic analysis in prior works. To our knowledge, this work provides among the first empirical characterizations of cross-dataset generalization behavior for HLS QoR models.

A clear asymmetry emerges between using synthetic vs. benchmark designs as training data sources. Training on synthetic designs, whether \textbf{\textit{Base}} or \textbf{\textit{Final}} sets, transfers well to testing on benchmark designs (green regions). \textit{The reverse does not hold.} Models trained on benchmark designs generalize poorly to testing on synthetic designs (red region). Notably, models trained on synthetic designs and evaluated on benchmark test sets also outperform models both trained and evaluated entirely on benchmark designs only (yellow region). These results imply that synthetic designs provide a richer and more diverse training set than designs from existing HLS benchmarks.

Between the two synthetic design sources, the \textbf{\textit{Final}} iteratively mutated designs show stronger generalization. The results show both training on \textbf{\textit{Base}} and training on \textbf{\textit{Final}} yield comparable performance when evaluated on benchmark designs, with latency prediction showing \textbf{\textit{Final}} designs slightly outperforming \textbf{\textit{Base}} designs (green regions). However, we also find that models trained on \textbf{\textit{Final}} generalize better to testing on \textbf{\textit{Base}} than training on \textbf{\textit{Base}} and testing on \textbf{\textit{Final}} (purple region). This second asymmetry implies that \textbf{\textit{Final}} designs are the most broadly transferable training corpus.

Taken as a whole, these findings suggest that synthetic iteratively mutated designs generated by \ourwork{} offer the most robust foundation for HLS QoR model training.

\subsection{\textbf{Synthetic Design Trajectories}}

To better understand the proposed feedback-guided design mutation flow, we quantify the ability of an LLM within \ourwork{} to mutate designs while preserving validity, as well as the ability of \ourwork{} to push targeted design metrics in the directions we indicate.

\textbf{Valid Design Pass Rates}: Fig.~\ref{fig:pass_rates} shows the ability of LLMs to generate valid designs, both at the seed design generation stage and at each subsequent design mutation stage. We report validity using the pass@k metric, where pass@k is the probability that at least one of $k$ sampled designs is valid (i.e., compiles, simulates correctly, and synthesizes). Looking at the average pass@8 rate across domains, we see that using the \texttt{gpt-oss-120B} LLM model, \ourwork{} generates at least one valid, synthesizable mutated design sample for each source design in almost all cases. Looking at the pass@1 rate, equivalent to the average pass rate among sampled mutated designs, we see that the average pass rate across mutation stages is between 30\% and 80\%, indicating that on average most designs yield more than one passing sample from which to rank and select the best mutation. Together, these results indicate that \ourwork{} generates at least one valid mutated sample per design, enabling progression through mutation stages, while typically producing multiple valid samples, providing the diversity needed to rank and select the best mutation to pass to the next stage.

The sole exception to the pass rate trends observed is the "Engineering Simulation" domain, whose designs never produce any valid mutated HLS designs after the first mutation stage. In practice, this stems from the LLM performing poorly at seed generation for this domain, producing only one valid HLS seed design; this single design's trajectory then ends after none of its mutated samples are valid. This outlier highlights that high-quality seed design generation is a critical aspect of the \ourwork{} flow.

\begin{figure}[t!]
    \centering
    \includegraphics[width=0.5\textwidth]{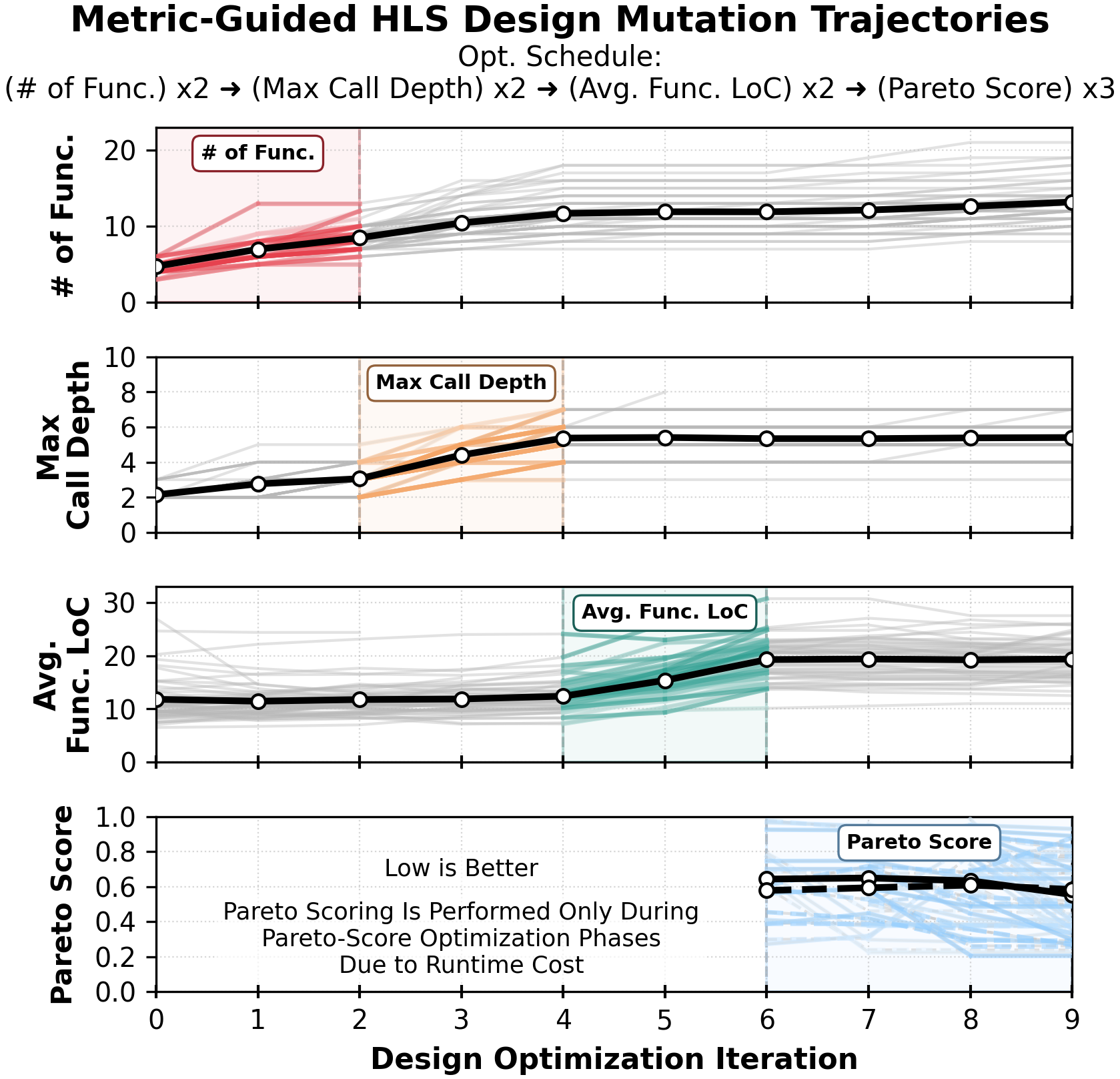}
    \caption{Mutation trajectories for all synthetic HLS designs from \ourwork{}. Thin lines track individual designs; bold shows the mean. Highlighted regions mark each metric's targeted optimization phase. Non-targeted metrics remain stable or improve passively alongside targeted ones.}
    \label{fig:trj}
\end{figure}

\textbf{Metric-Guided Mutation Trajectories}: Fig.~\ref{fig:trj} illustrates the metric-guided mutation trajectories of all HLS designs from a design generation run of \ourwork{} under a specified optimization schedule. During the optimization iterations, the scheduled target metrics generally improve in the intended direction: the code complexity metrics increase and Pareto scores decrease. At the same time, non-target metrics generally remain stable or continue to improve without substantial regression. Overall, these results show \ourwork{} achieves its designed goal to diversify designs along different dimensions through guided iterative mutation.

We note that the mutation schedule driving these trajectories is user-specified. In our schedule, we order structural objectives before Pareto-score objectives, as design-space scalability is most meaningful once a design has sufficient structure to parameterize. Similarly, the non-target metric stability observed in Fig.~\ref{fig:trj} is encouraged at the prompt level rather than enforced as a hard constraint. Exploring dynamic or learned mutation schedules is a promising extension of this mutation flow.

\subsection{\textbf{Synthetic Design Dataset Characterization}}

\begin{figure}[t]
    \centering
    \includegraphics[width=\columnwidth]{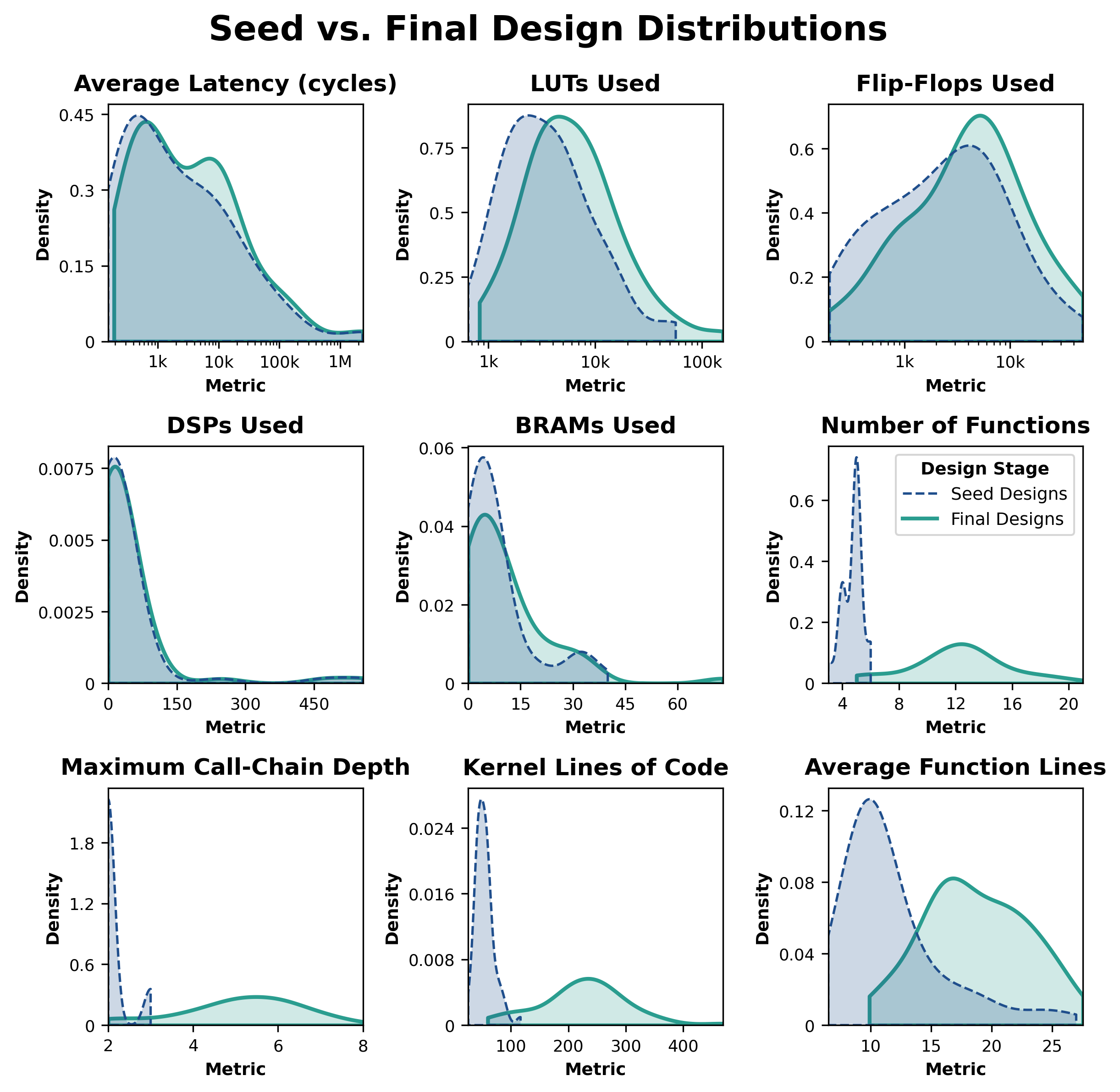}
    \caption{Distributions of nine design metrics for synthetic seed designs versus final mutated designs generated by \ourwork{}, covering post-synthesis QoR (latency and resource usage) and program structure (e.g., number of functions, call depth, function LoC). Final mutated designs exhibit notably more diverse program structure and slightly more diverse design metrics than seed designs.}
    \label{fig:dist}
\end{figure}
\begin{figure}[t]
    \centering
    \includegraphics[width=\columnwidth]{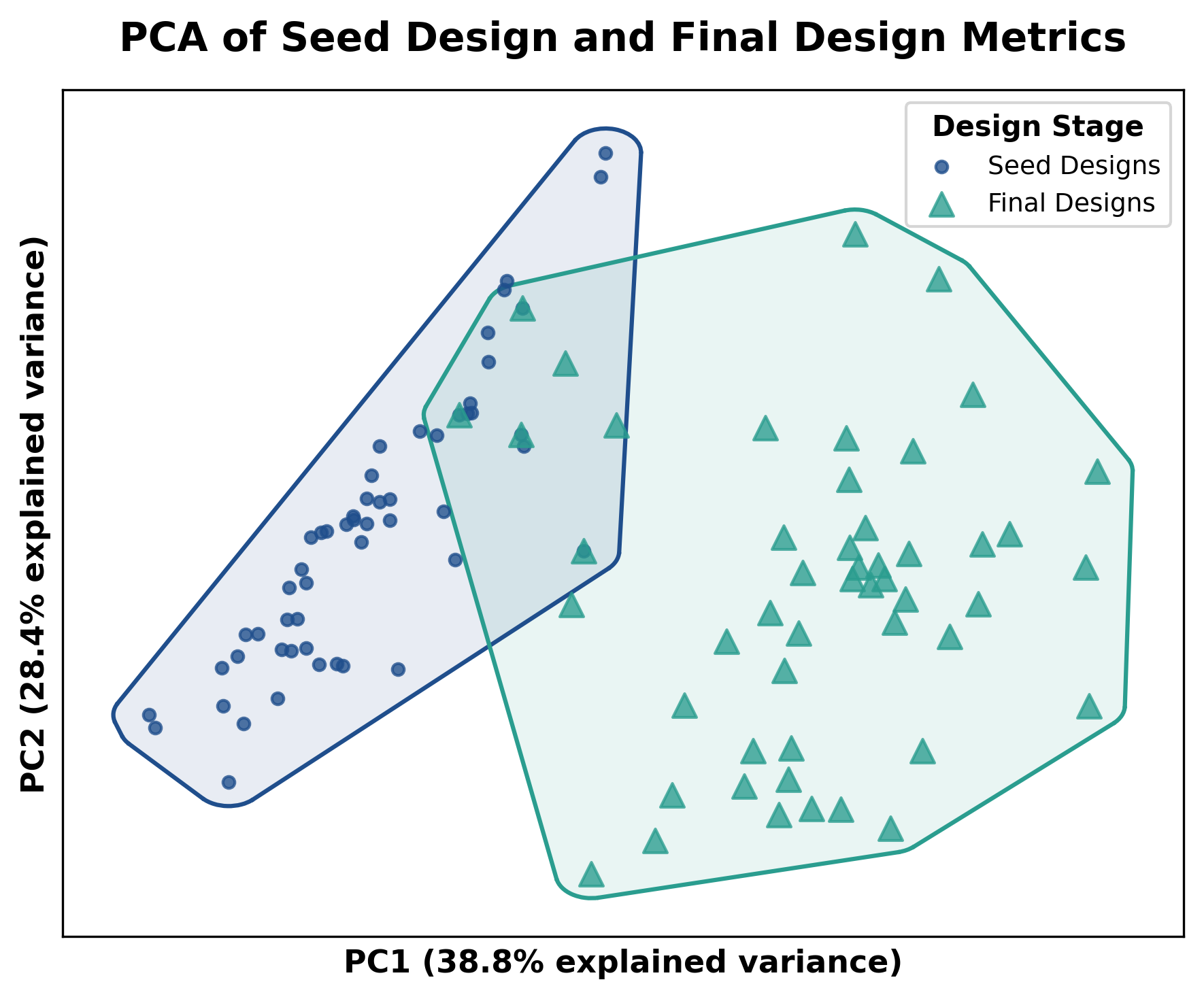}
    \caption{2D PCA projection of seed and final mutated designs, where each point represents one design described by its nine-metric feature vector (the same metrics as in Fig.~\ref{fig:dist}). The seed design space is narrow, restricted largely along a single direction, while the final mutated design space is broader and more uniform, indicating greater overall design diversity.}
    \label{fig:pca}
\end{figure}

To characterize both the synthetic seed designs and the final mutated designs generated by \ourwork{}, we examine the distributions of individual design metrics (Fig.~\ref{fig:dist}) and the joint diversity of designs across all metrics via dimensionality reduction (Fig.~\ref{fig:pca}).

\textbf{Individual Design Metrics}: From the distributions of design metrics in Fig.~\ref{fig:dist}, we make two observations: 1)~final mutated synthetic designs have a more diverse distribution of program structure than synthetic seed designs, and 2)~final mutated synthetic designs have slightly more diverse distributions of design metrics (latency and resource usage) than seed designs. Both observations strengthen the claim that mutated synthetically generated HLS designs are better than zero-shot generated designs.

Based on these distributions, we hypothesize that the greater diversity in program structure of mutated synthetic designs, rather than their diversity in design metrics (which is only marginally higher), is what enables better generalization for QoR proxy modeling. Further evaluation is needed to understand these generalization dynamics for HLS designs and to consider aspects not captured in these distributions, such as design space size and quality via Pareto metrics.

\textbf{Overall Design Space}: We characterize the diversity of the seed and final mutated synthetic designs across all design metrics jointly. Fig.~\ref{fig:pca} shows a dimensionality-reduction plot in which each point represents a design, described by a feature vector containing the same nine metrics shown in Fig.~\ref{fig:dist}. These design vectors are projected to two dimensions using PCA, revealing the "design space" of the synthetic seed designs and the final mutated designs. Qualitatively, the seed design space is far less diverse, restricted largely along a one-dimensional feature direction, while the final mutated design space is much larger and more uniform. This further supports our hypothesis that the broader design space of mutated synthetic designs leads to better generalization on deep learning tasks for hardware design.

These synthetic design characterization results raise a natural question: could the benefit of iterative mutation be matched by allocating equivalent inference compute to zero-shot generation, e.g., zero-shot seed design sampling with N equal to the total number of LLM calls invoked by iterative mutation? We argue this is unlikely: drawing more zero-shot samples does not imply greater design diversity in terms of size, complexity, or design-space quality, as samples likely remain bound to the distribution of the model's unconditioned zero-shot generations; feedback-guided mutation is what we hypothesize pushes designs beyond this distribution. Fig.~\ref{fig:pca} is consistent with this view: seed designs concentrate along a low-dimensional manifold, while mutated designs are able to move off this manifold into a broader design space.

\section{Conclusion}

We presented \ourwork{}, a framework for generating diverse, synthesizable HLS design datasets via iterative LLM-guided mutation. Through cross-validation, we demonstrated that our iteratively mutated synthetic designs provide the most generalizable training corpus for HLS QoR modeling among existing benchmark designs and zero-shot designs, and our design characterization analyses show that feedback-guided mutation pushes designs into a broader and more diverse design space than zero-shot generation alone can reach.

Since \ourwork{} scales corpus size directly with LLM and synthesis compute budget, this opens several natural directions for future work: generation across a broader range of LLMs, rigorous compute-matched and inference-scaling comparisons against zero-shot baselines, adaptive mutation schedules, and agentic design mutation.

We hope that by open-sourcing \ourwork{}, we enable others to build on our framework, extending it with new mutation strategies, design metrics, and LLM evaluations. In doing so, we hope to drive the curation of larger and more diverse HLS design datasets for the broader hardware design community.

\bibliographystyle{IEEEtran}
\bibliography{refs}

\end{document}